# $[Ag(NH_3)_2]_2SO_4$: A coordination strategy on the cationic-moiety to design nonlinear optical materials

Yi-Chang Yang, Xin Liu, Jing Lu, Li-Ming Wu* and Ling Chen*

[*] Y.-C. Yang, Prof. Dr. L. Chen
Beijing Key Laboratory of Energy Conversion and Storage Materials
College of Chemistry, Beijing Normal University
Beijing 100875 (P. R. China)
E-mail: chenl@bnu.edu.cn
Ling Chen ORCID: 0000-0002-3693-4193

X. Liu, J. Lu, Prof. L.-M. Wu
Key Laboratory of Theoretical and Computational Photochemistry
Ministry of Education, College of Chemistry
Beijing Normal University, Beijing 100875 (P. R. China)
E-mail: wlm@bnu.edu.cn
Li-Ming Wu ORCID: 0000-0001-8390-2138

Supporting information for this article is given via a link at the end of the document.

**Abstract:** Over decades, guided by the anionic group theory, the majority work has been focused on the anionic-moiety of a nonlinear optical material, however, the property guided structure modification and design on the cationic-moiety has long been neglected. Herein, we report for the first time a coordination strategy on the cationic-moiety, as demonstrated by the first exmple, $[Ag(NH_3)_2]_2SO_4$ vs $Ag_2SO_4$, the coordination of the plus one $Ag^+$ cation by the neutral ligand forming the $[Ag(NH_3)_2]^+$ cationic-moiety drives the formation of the noncentrosymmetric tetragonal $P\bar{4}2_1c$ structure which exhibits a remarkable property improvement, including a strong SHG intensity (1.4 × KDP vs 0 @1064 nm), and a large birefringence ($\Delta n_{cal.}$: 0.102 vs 0.012 @1064 nm). Furthermore, we discover that owing to the strong hydrogen bonds and spatial confinement of $[SO_4]^{2-}$ anions, the cation $[Ag(NH_3)_2]^+$ bends parallel to the crystallographic *c* axis with a bond angle of N-Ag-N = 174.35°, generating a permenant dipole moment $\mu_z$ = -0.12 D that is responsible for the large birefringence that enlarges the range of the phase matching SHG laser output. We believe what we discover is only a glimpse of the long been neglected crystal engineering on the cationic-moiety. Encourage by this work, exciting works on other ionic compound systems shall flood in.

Nonlinear optical (NLO) materials have attracted intense attentions recently driven by the fast growing demands of array of fascinating applications, such as precision manufacturing, information processing and medical treatment, etc.[1] The most important and the majority of the widely applied NLO materials are inorganic NLO materials, some of them are successfully commercialized, such as KBBF,[2] $KH_2PO_4$ (KDP),[3] $\beta$-$BaB_2O_4$ (BBO),[4] and $LiB_3O_5$ (LBO),[5] etc.[6-8] To meet the real application needs, these materials shall satisfied the comprehensive requirements, including large second harmonic generation (SHG) coefficient, suitable band gap ($E_g$), appropriate birefringence ($\Delta n$) and good physical and chemical stability. However, to have all these properties at once in a single structure is not always possible, because the SHG intensity is somewhat inversely correlated with the $E_g$; whereas the $\Delta n$ determined by the optical anisotropy, is tangled with the SHG and $E_g$ complicatedly. Especially, for the material constructed by a non-π conjugated anionic group, the birefringence is usually pretty small owing to the isotropic geometry nature of the building unit, which eventually limits the phase matching SHG output.

As we know, for an ionic NLO material, the structure contains the cationic and anionic structure moieties. According to the anionic group theory,[9] the SHG is mainly contributed from the anionic-moiety, as demonstrated in KBBF,[2] KDP,[3] BBO,[4] and LBO,[5] etc. In particularly, the element identity of the anionic group, such as $[PO_4]^{3-}$ vs $[PO_3F]^{2-}$, and the inter-group connection and aggregation modes, as well as the orientation of the anionic groups are all crucial to influence the SHG, which constituents the majority of the research works.[10-13] Regarding sulfate, it is newly recognized as a NLO material in the UV region, for instance, $Li_8NaRb_3(SO_4)_6{\cdot}2H_2O$, (SHG intensity: 0.5 × KDP)[14] $NH_4NaLi_2(SO_4)_2$ (1.1 × KDP)[15] and $CsSbF_2SO_4$, (3.0 × KDP).[16]

However, the influence of the cationic-moiety on improving the NLO property has long been neglected. None reported is found to date. One may argue that cations carrying second-order Jahn-Teller effect have enhanced the SHG intensity by causing a great coordination sphere distortion,[16-17] but these cations construct the anionic-moiety, instead. For example, in $CsSbF_2SO_4$, the anionic-moiety is constructed by $[Sb^{3+}O_4F_2]^{7-}$ and $[SO_4]^{2-}$.[16] The anionic-moiety of KTP is made up by $[TiO_6]^{8-}$ and $[PO_4]^{3-}$. Therefore, the crystal engineering on the cationic-moiety to improve the property remains untouched. Considering the high electropositivity of the alkali or alkali earth metal cations in the conventional NLO materials, the cationic-moiety remains as simple as an array of spherical counter cations to balance the negative charge of the complicated anionic skeleton. That is why people think there is no need to study to cationic-moiety of an NLO material.

In this work, we present for the first time a coordination strategy on the cationic-moiety, which successfully turns the SHG inert orthorhombic $Ag_2SO_4$ structure into an NLO active $[Ag(NH_3)_2]_2SO_4$ structure. The $Ag^+$ cation is linearly coordinated by two neutral ammonia ligands, which not only drives the formation of the noncentrosymmetric structure exhibiting a strong phase matching SHG intensity (1.4 × KDP), but also enhances greatly the birefringence (0.102 vs 0.012 of $Ag_2SO_4$ @1064 nm), which facilitates the phase matching SHG laser output ($\lambda_{PM}$) down to 281 nm. We believe this discover is only a glimpse of the iceberg of the long been neglected crystal engineering on the cationic-moiety. With such a beam of useful light, a new direction of the efforts to improve the performance and design and searching of NLO materials shall encourage exciting works flood in.

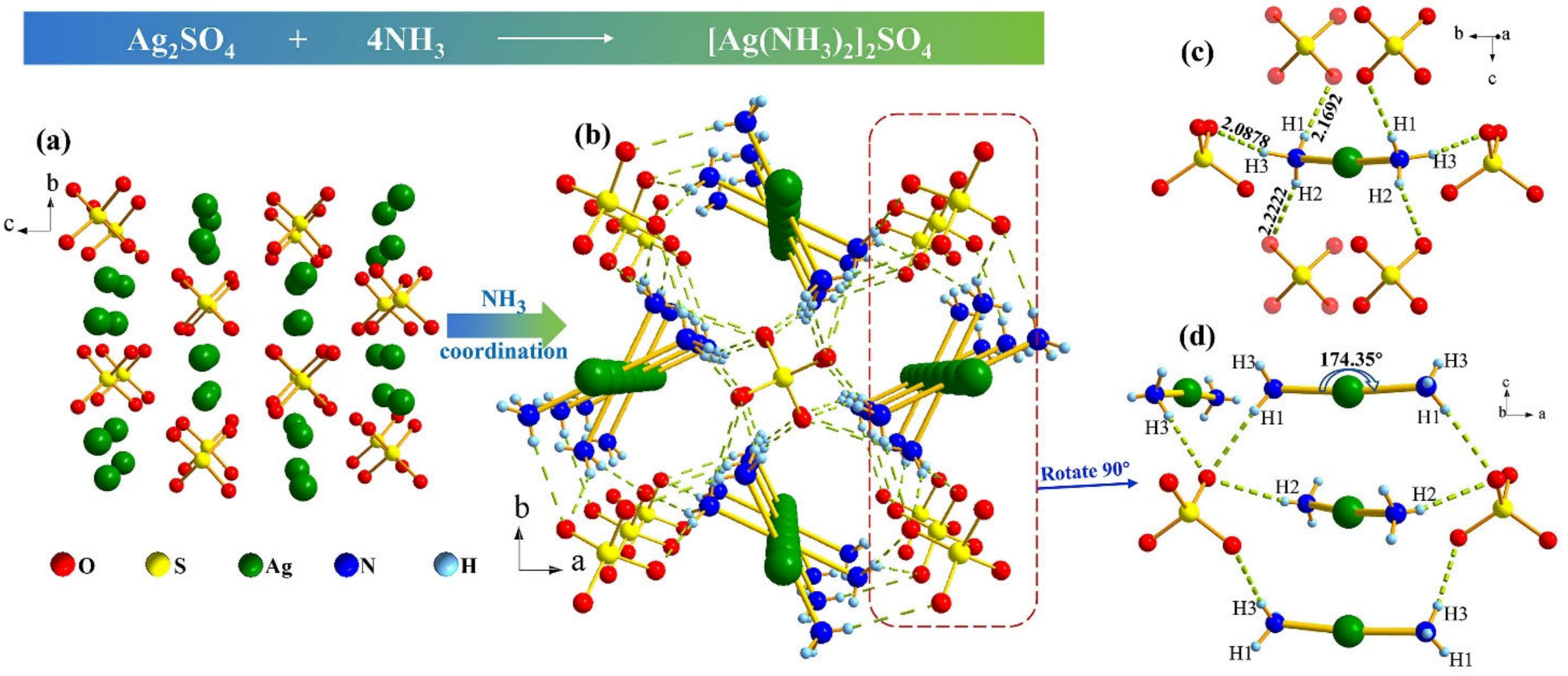

**Figure 1.** Perspective views of crystal structures of (a) $Ag_2SO_4$ along [100] and (b) $[Ag(NH_3)_2]_2SO_4$ along [001]. Green dashed lines: Hydrogen bonds. (c) Hydrogen bonding environment around the $[Ag(NH_3)_2]^+$ cation. (d) Bent structure of $[Ag(NH_3)_2]^+$ with a bond angle of N-Ag-N = 174.35° parallel to the crystallographic *c* axis.

Colourless crystals of $[Ag(NH_3)_2]_2SO_4$ with sizes up to 2.8 × 1.4 × 0.4 cm$^3$ (Figure 2a inset) were successfully obtained by a modified method of Jacobs[18]. Different from the reported instability of this compound, we find $[Ag(NH_3)_2]_2SO_4$ is stable under the ambient condition for more than six months, moreover it undergoes a 1$^{st}$ order phase transition at 83 °C. (Figure S3) Besides, we can grow crystals with sizes roughly 10 times larger than the reported.[18] The property measurement hereafter is carried out at room temperature unless it is noted to ensure the fine mono-phase purity. The phase purity was confirmed by the well indexed powder XRD pattern (Figure 2a) and the successful assigned characteristic IR vibrations (Figure S2 and Table S5). Our single crystal X-ray diffraction data confirm the $[Ag(NH_3)_2]_2SO_4$ a tetragonal $P\bar{4}2_1c$ (No. 114) structure with $a = b$ = 8.4372 (10) Å, and $c$ = 6.3980(8) Å (Table S1) in good agreement with the previous report.[18] (Figure 1a,b) This structure shows a very nice structural change from the centrosymmetric orthorhombic *Fddd* (No. 70) $Ag_2SO_4$ structure, mostly caused by the coordination of the plus one cation $Ag^+$ to the neutral ammonia ligands. (Figure 1d) The coordinated cationic-moiety features a silver-over-silver stacks of the $[Ag(NH_3)_2]^+$ cations along the crystallographic *c* axis with a Ag-Ag separation of 3.1990 Å. Each $Ag^+$ is coordinated by two neutral ammonia molecules, forming a strongly anisotropic $[Ag(NH_3)_2]^+$ linear dumbbell geometry with a bond length of Ag-N = 2.1072 Å and each ammonia forms three hydrogen bonds with distance of N-H1···O = 2.1692 Å; N-H2···O = 2.2222 Å; and N-H3···O = 2.0878 Å, respectively. (Figure 1c) In the mother $Ag_2SO_4$ structure, the $Ag^+$ cation is rather an isolated sphere plus one cation showing a $O_h$ $[Ag^+O_6]$ geometry without any significant covalent Ag-O bonding[19] interaction. In comparison, the anisotropic linear $[Ag(NH_3)_2]^+$ dumbbell cation is held together by strong covalent Ag-N bonds, and further as a whole, is linked to a stable three-dimensional network via strong hydrogen N-H···O bonds with the adjacent anionic arrays of isolated $[SO_4]^{2-}$. Driven by such an ammonia coordination interaction, the SHG inert $Ag_2SO_4$ structure has been changed into an SHG active noncentrosymmetric $[Ag(NH_3)_2]_2SO_4$ structure.

The anionic-moiety in $[Ag(NH_3)_2]_2SO_4$, on the other hand, is a simple array of isolated $[SO_4]^{2-}$ anions, which undergoes nearly a neglectable distortion with S-O bond of 1.4694 Å, and O-S-O angles of 109.9(2) or 109.25(10)°. Therefore, the isotropic tetrahedral $[SO_4]^{2-}$ anion contributes only trivial to the anisotropy of the material. Note that, owing to the linkage of the hydrogen bonds as well as the spatial confinement of the adjacent arrays of $[SO_4]^{2-}$ anions, the $[Ag(NH_3)_2]^+$ dumbbell bends towards the crystallographic *c* axis with a bond angle of N-Ag-N = 174.35°. (Figure 1d) Interestingly, the N-Ag-N triangle plane parallels to the *c* axis. Such a bent $[Ag(NH_3)_2]^+$ dumbbell obeys the operation of the rotoinversion axis $\bar{4}$, through which the difference on the crystallographic *ab* plane caused by its nature of the linear geometry has been diminished; meanwhile, the $P\bar{4}2_1c$ symmetry requirement is satisfied. According to the linear optical classification, the tetragonal crystal system is classified as a uniaxial crystal, therefore the ordinary index $n_o$ lies in the crystallographic (001) plane. Remarkably, the N-Ag-N = 174.35° bending gives rise to a permanent static dipole moment of $\mu_z$ = -0.12 D (Table S4). As the refractive indices ($n$) is proportional to the strength of dipole moment, $n \propto \mu$, when the light propagates through such an induced electric field generated by the permanent dipole, the propagation speed is slowed down, giving a larger $n_e$. Consequently, confined by such a structure feature, we can deduce that $[Ag(NH_3)_2]_2SO_4$ shall be a positive uniaxial crystal. This statement is verfied by the birefringence measurement as well as the DFT calculations[20] on the refractive index dispersion curve discussed below.

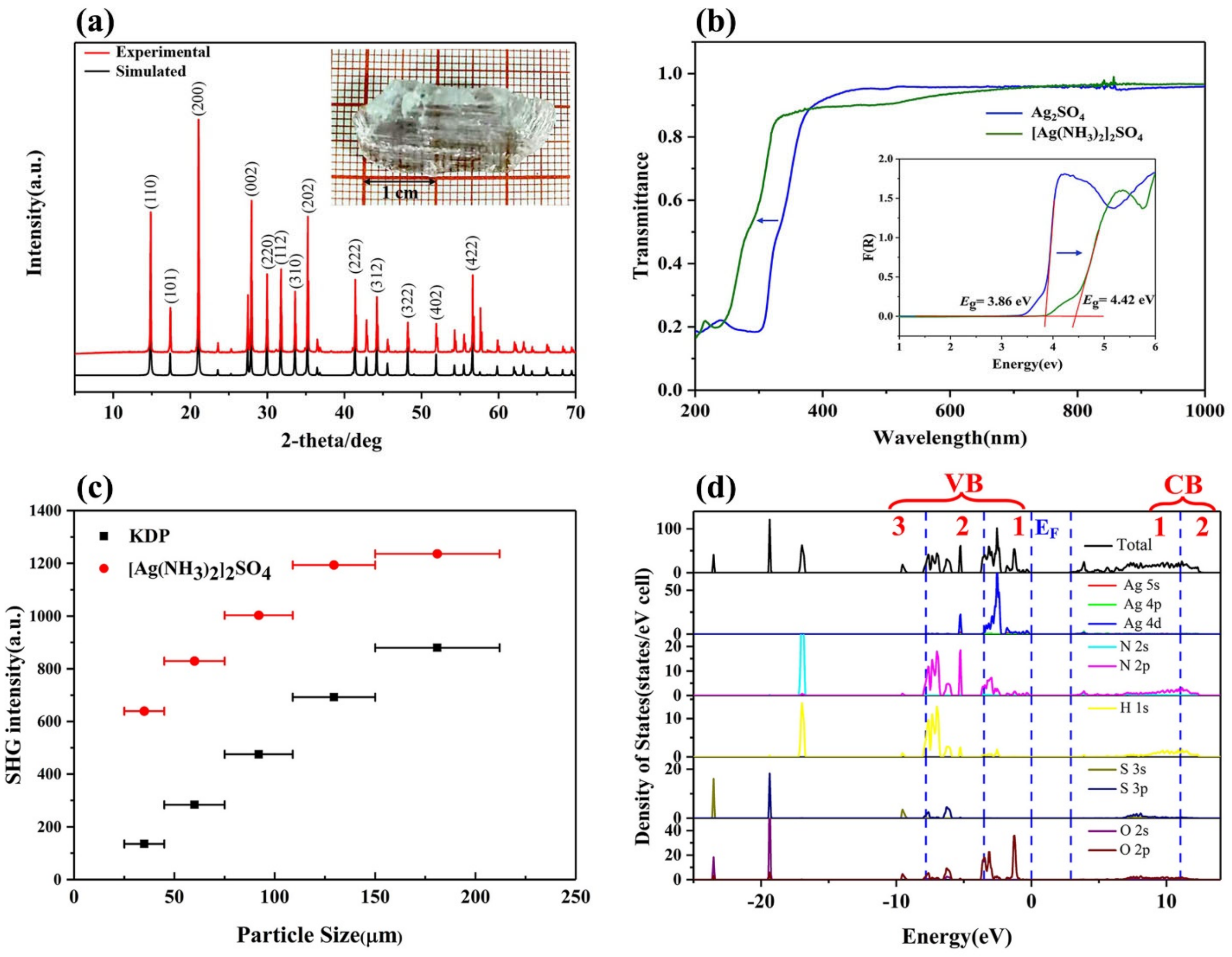

**Figure 2.** (a) Experimental and simulated PXRD patterns of $[Ag(NH_3)_2]_2SO_4$. Inset: Photo of the centimeter-sized as-grown crystal of $[Ag(NH_3)_2]_2SO_4$. (b) UV-vis-NIR diffuse reflectance spectra of polycrystalline samples of $Ag_2SO_4$ and $[Ag(NH_3)_2]_2SO_4$. Inset: Optical diffuse reflectance spectra. (c) Powder SHG intensity curves at 1064 nm. (d) Total and partial densities of states of $[Ag(NH_3)_2]_2SO_4$.

The UV-vis-NIR diffuse reflectance spectra (Figure 2b) indicate $Ag_2SO_4$ and $[Ag(NH_3)_2]_2SO_4$ a wide transparency window ranging from the near-UV to the near-IR region with a UV cutoff absorption edge of 321 and 281 nm, respectively, which gives a band gap ($E_g$) of 3.86 and 4.42 eV. These values agree with those calculated, 1.25 (Figure S6a) and 2.93 eV (Figure S5a), respectively. The band structure and density of state analyses reveal the $E_g$ enhancement in $[Ag(NH_3)_2]_2SO_4$ is attributed to the coordination of ammonia to the $Ag^+$ cation, which pushes the Ag 5*s* states which had constituted, as a major component, the conduction band minimum (CBM) of the mother $Ag_2SO_4$ structure (Figure S6b), up to the higher energy levels. As a consequence, the band gap widens as shown in Figure 2b, in which the CBM is now mainly from the Ag 4*d* and N 2*p* states.

As shown in Figure 2c, $[Ag(NH_3)_2]_2SO_4$ exhibits a phase matching behavior with SHG efficiency maximized at an intensity that is about 1.4 times that of KDP ($KH_2PO_4$). This intensity is the strongest among sulfates to date, if $CsSbF_2SO_4$ (3.0 × KDP) [16] and $(NH_4)SbCl_2SO_4$ (1.7 × KDP),[21] having the $Sb^{3+}$ cation with Jahn-Teller effect are excluded. And the SHG intensities of other sulfates are weaker, such as $Li_8NaRb_3(SO_4)_6{\cdot}2H_2O$, (0.5 × KDP)[14] $NH_4NaLi_2(SO_4)_2$ (1.1 × KDP).[15]

In more details, the theoretical calculations of the band structure and total and partial density of states (DOS) based on the DFT method[20] tell the $E_g$ of $Ag_2SO_4$ (Figure S6b) is mainly determined by the Ag 4*d* and O 2*p* in the valence band maximum (VBM), and the Ag 5*s* in the CBM. In comparison, the VBM of $[Ag(NH_3)_2]_2SO_4$ (Figure 2d) contains the N 2*p* except for the Ag 4*d* and O 2*p*. And the latter constitute the VBM of the mother compound. Meanwhile, the CBM now mainly comes from Ag 4*d* and N 2*p*, instead of the Ag 5s only in the mother compound. (Figure S7) Clearly, during the $Ag^+$ coordination process to $NH_3$, the 5*s* and 4*p* orbitals of the Ag atom undergo a linear sp-hybridization to accommodate the electrons donated from N atom of the ammonia molecule, consequently, the Ag 5*s* state is pushed up to the higher energy level. As a result, the cut-off absorption edge of $[Ag(NH_3)_2]_2SO_4$ is 40 nm shorter than that of $Ag_2SO_4$.

The cutoff-energy-dependent SHG coefficient[22] analyses according to the length-gauge formalism,[23] as shown in Figure S5b, reveal the SHG response originates dominantly from the VB-1 and CB-1 regions, thus, the states dominating the VB-1 region of Ag 4*d* and O 2*p* mixing with minor N 2*p* as well as those unoccupied Ag 4*d* and N 2*p* states that predominate the CB-1 region are responsible mainly for the SHG effect, this means both the $[SO_4]^{2-}$ anionic-moiety and the $[Ag(NH_3)_2]^+$ cationic-moiety contribute to the SHG. The calculated $d_{36}$ of $[Ag(NH_3)_2]_2SO_4$ and KDP are 1.50 vs 1.1 pm/V at 1064 nm, (Figure S8) which in principle agree with the experimental observation.

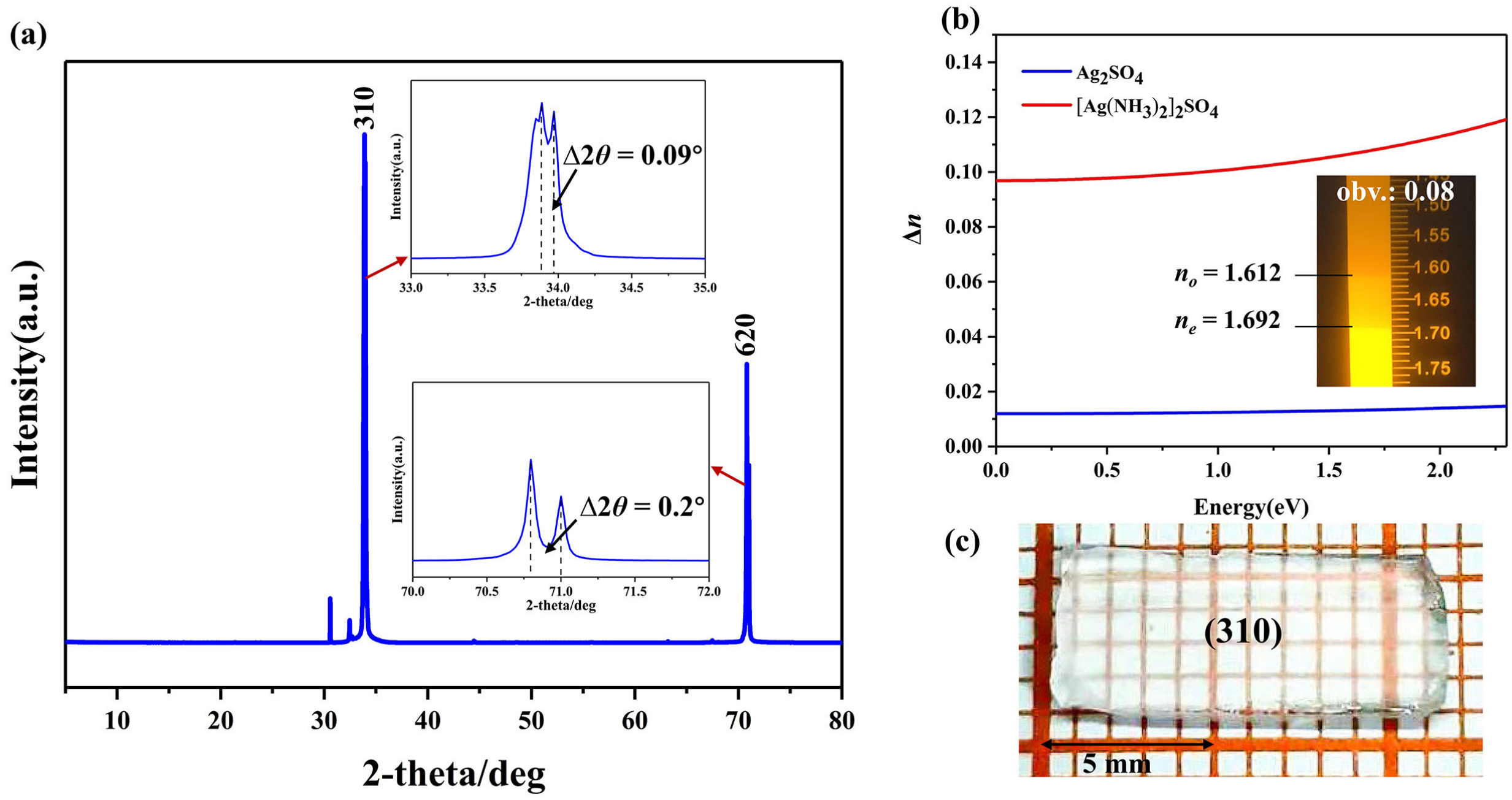

**Figure 3.** (a) XRD patterns of the hand-polished $[Ag(NH_3)_2]_2SO_4$ crystal slice, the doublets around 34° and 71° are due to the difference of the Cu *Kα*1 and *Kα*2. (b) The calculated birefringences of $Ag_2SO_4$ and $[Ag(NH_3)_2]_2SO_4$. Inset: Photo of the birefringence measured by a hand-polished single crystal slice of $[Ag(NH_3)_2]_2SO_4$. (c) Hand-polished crystal of $[Ag(NH_3)_2]_2SO_4$ used to measure the properties in this figure.

The birefringence ($\Delta n$) has been measured on a hand-polished single crystal slice of $[Ag(NH_3)_2]_2SO_4$ (Figure 3c) utilizing the immersion technique[24] at the wavelength of 589.3 nm. As Figure 3a indicated, such a hand-polished crystal slice exhibits the profound (310) (620) diffractions, which means the crystallographic *c* axis lies in the plane parallel to the front surface of the slice. As shown in Figure 3b inset and Table S6, we observe the minimum $n_o$ = 1.605, and as we rotate the slice clockwise, the $n_e$ varies a lot, and maximizes at 1.693. As a result, a $\Delta n_{obv.}$ = 0.088 is obtained at the incident light of 589.3 nm. Since the crystal is not perfectly cut and polished, the actual $\Delta n$ value should be larger than this value, which is supported by the $\Delta n_{cal.}$ = 0.115 at the same energy. (Figure 3b) The calculation also support that $[Ag(NH_3)_2]_2SO_4$ is a positive uniaxial crystal with $n_e > n_o$. (Figure S10) Besides, the birefringence of $[Ag(NH_3)_2]_2SO_4$ ($\Delta n_{cal.}$ = 0.102@1064 nm) is significantly larger than that of the mother compound $Ag_2SO_4$ (0.012@1064 nm). The small $\Delta n$ of $Ag_2SO_4$ is owing to its optical isotropic nature that is determined mainly by the structural feature of which the cationic-moiety is an isotropic $Ag^+$ spherical cation, whereas the anionic-moiety is an isotropic $[SO_4]^{2-}$ tetrahedron. Apparently, the coordination of the cationic-moiety in $[Ag(NH_3)_2]_2SO_4$ greatly enhances the birefringence by roughly one order of magnitude. (Figure 3b) To further understand the source of birefringence, the charge density of the two moieties in the $[Ag(NH_3)_2]_2SO_4$ structure are studied. As listed in Table S4, the polarizability anisotropy of $[Ag(NH_3)_2]^+$ (12.84) overwhelms that of $[SO_4]^{2-}$ (0.0093). Regarding the static dipole moment, those of $[SO_4]^{2-}$ are almost zero along the *x*, *y* and *z* axes with -0.00051, 0, and 0.0020 D, respectively. Differently, as the 174.35° bending of $[Ag(NH_3)_2]^+$ generates a static dipole moment ($\mu$) of -0.12 D along the *z* axis, which is 60 times larger than that of $[SO_4]^{2-}$ anion. These data indicate that the $[Ag(NH_3)_2]^+$ cationic-moiety gives rise to the large birefringence. Since $\mu_z > \mu_x, \mu_y$, thus, the light propagation speed is slower in the *c* direction, leading to a larger $n_e$ than the $n_o$ in the *ab* plane, i.e., $n_e > n_o$, consequently, the tetragonal $[Ag(NH_3)_2]_2SO_4$ is a positive uniaxial crystal.

In summary, we report herein for the first time a coordination strategy on the cationic-moiety, which leads to the discovery of $[Ag(NH_3)_2]_2SO_4$, in which the remarkable enhancement of nonlinear and linear optical properties are owing to the unique cationic-moiety structure. We demonstrate the coordination of the electrical neutral ammonia molecules to the spherical $Ag^+$ cation drives the formation of the noncentrosymmetric tetragonal structure, widens the energy gap by pushing up the Ag 5*s* states through the strong coordination interactions. Remarkably, the linkage of the strong hydrogen bonds and the spatial confinement of the adjacent arrays of isolated $[SO_4]^{2-}$ anions bend the $[Ag(NH_3)_2]^+$ cation parallel to the crystallographic *c* direction with a bond angle of N-Ag-N = 174.35°, which generates a permenant static dipolement $\mu_z$ that leads to $n_e > n_o$, and eventually giving rise to the positive uniaxial crystal nature. In comparison with the mother compound $Ag_2SO_4$, the tetragonal $[Ag(NH_3)_2]_2SO_4$ exhibits a large birefringence ($\Delta n_{cal.}$ = 0.102 vs 0.012), and a strong SHG response (1.4 × KDP vs 0, which agrees with the calculated $d_{36}$ = 1.50 pm/V). The $\Delta n$ is observed to be 0.088 at 589.3 nm on a hand-polished (310) crystal slice. Theoretical studies also confirm that the $[Ag(NH_3)_2]^+$ cationic-moiety is responsible mainly for the optical anisotropy and contributes importantly to the SHG effect. Moreover, this compound shows a feasibility of large size crystal growth. We believe what we discover herein is only a glimpse of the long been neglected of the crystal engineering on the cationic moiety of a structure. With such a beam of useful light, this new direction to property improvement, rational design and searching of high-performance function materials, shall encourage floods of excellent works.

## Acknowledgements

This research was supported by the National Natural Science Foundation of China (21971019) and by the Beijing Natural Science Foundation (2202022).

### Conflict of interest

The authors declare no conflict of interest.

[1] a) D. Cyranoski, *Nature* **2009**, *457*, 953-955; b) U. Gubler, C. Bosshard, *Nat. Mater.* **2002**, *1*, 209-210; c) X. T. Wu, L. Chen, in *Structure-Property Relationships in Non-Linear Optical Crystals I: The Uv-Vis Region, Vol. 144* (Eds.: X. T. Wu, L. Chen), **2012**, pp. XI-XII.

[2] C. T. Chen, G. L. Wang, X. Y. Wang, Z. Y. Xu, *Appl. Phys. B: Lasers Opt.* **2009**, *97*, 9-25.

[3] D. M. Burland, *Chem. Rev.* **1994**, *94*, 1-2.

[4] C. T. Chen, B. C. Wu, A. D. Jiang, G. M. You, *Sci. Sin., Ser. B (Engl. Ed.)* **1985**, *28*, 235–243.

[5] C. T. Chen, Y. C. Wu, A. D. Jiang, B. C. Wu, G. M. You, R. K. Li, S. J. Lin, *J Opt Soc Am B* **1989**, *6*, 616-621.

[6] H.-M. Zhou, L. Xiong, L. Chen, L.-M. Wu, *Angew. Chem. Int. Ed.* **2019**, *58*, 9979-9983.

[7] R. C. Eckardt, Y. X. Fan, R. L. Byer, R. K. Route, R. S. Feigelson, J. Vanderlaan, *Appl. Phys. Lett.* **1985**, *47*, 786-788.

[8] G. D. Boyd, E. Buehler, F. G. Storz, *Appl. Phys. Lett.* **1971**, *18*, 301-302.

[9] a) C. T. Chen, Y. B. Wang, B. C. Wu, K. C. Wu, W. L. Zeng, L. H. Yu, *Nature* **1995**, *373*, 322-324; b) C. T. Chen, *Sci. Sinica* **1979**, *22*, 756-776.

[10] a) C. T. Chen, Y. C. Wu, R. K. Li, *Int. Rev. Phys. Chem.* **1989**, *8*, 65-91; b) K. M. Ok, *Acc. Chem. Res.* **2016**, *49*, 2774-2785; c) L. Kang, F. Liang, X. Jiang, Z. Lin, C. Chen, *Acc. Chem. Res.* **2020**, *53*, 209-217.

[11] P. Yu, L. M. Wu, L. J. Zhou, L. Chen, *J. Am. Chem. Soc.* **2014**, *136*, 480-487.

[12] S. G. Zhao, P. F. Gong, S. Y. Luo, L. Bai, Z. S. Lin, C. M. Ji, T. L. Chen, M. C. Hong, J. H. Luo, *J. Am. Chem. Soc.* **2014**, *136*, 8560-8563.

[13] a) L. Li, Y. Wang, B. Lei, S. Han, Z. Yang, K. R. Poeppelmeier, S. Pan, *J. Am. Chem. Soc.* **2016**, *138*, 9101-9104; b) Y. G. Shen, Y. Yang, S. G. Zhao, B. Q. Zhao, Z. S. Lin, C. M. Ji, L. Li, P. Fu, M. C. Hong, J. H. Luo, *Chem. Mater.* **2016**, *28*, 7110-7116.

[14] Y. Li, S. Zhao, P. Shan, X. Li, Q. Ding, S. Liu, Z. Wu, S. Wang, L. Li, J. Luo, *J. Mater. Chem. C* **2018**, *6*, 12240-12244.

[15] Y. Li, F. Liang, S. Zhao, L. Li, Z. Wu, Q. Ding, S. Liu, Z. Lin, M. Hong, J. Luo, *J. Am. Chem. Soc.* **2019**, *141*, 3833-3837.

[16] X. Dong, L. Huang, C. Hu, H. Zeng, Z. Lin, X. Wang, K. M. Ok, G. Zou, *Angew. Chem. Int. Ed.* **2019**, *58*, 6528-6534.

[17] a) P. S. Halasyamani, K. R. Poeppelmeier, *Chem. Mater.* **1998**, *10*, 2753-2769; b) F. F. Mao, C. L. Hu, X. Xu, D. Yan, B. P. Yang, J. G. Mao, *Angew. Chem. Int. Ed.* **2017**, *56*, 10630-10630; c) G. Zou, C. Lin, H. Jo, G. Nam, T. S. You, K. M. Ok, *Angew. Chem. Int. Ed.* **2016**, *55*, 12078-12082.

[18] U. Zachwieja, H. Jacobs, *Z. Kristallogr.* **1992**, *201*, 207-212.

[19] N. E. Brese, M. Okeeffe, B. L. Ramakrishna, R. B. Vondreele, *J. Solid State Chem.* **1990**, *89*, 184-190.

[20] W. Kohn, *Rev. Mod. Phys.* **1999**, *71*, 1253-1266.

[21] F. He, Q. Wang, C. Hu, W. He, X. Luo, L. Huang, D. Gao, J. Bi, X. Wang, G. Zou, *Cryst. Growth Des.* **2018**, *18*, 6239-6247.

[22] a) Y. Huang, L. Wu, X. Wu, L. Li, L. Chen, Y. Zhang, *J. Am. Chem. Soc.* **2010**, *132*, 12788-12789; b) M. H. Lee, C. H. Yang, J. H. Jan, *Phys. Rev. B* **2004**, *70*.

[23] a) C. Aversa, J. E. Sipe, *Phys. Rev. B* **1995**, *52*, 14636-14645; b) S. N. Rashkeev, W. R. L. Lambrecht, B. Segall, *Phys. Rev. B* **1998**, *57*, 3905-3919.

[24] C. T. Chen, Y. B. Wang, Y. N. Xia, B. C. Wu, D. Y. Tang, K. C. Wu, W. R. Zeng, L. H. Yu, L. F. Mei, *J. Appl. Phys.* **1995**, *77*, 2268-2272.

# Table of Contents

## $[Ag(NH_3)_2]_2SO_4$: A coordination strategy on the cationic-moiety to design nonlinear optical materials

Yi-Chang Yang, Xin Liu, Jing-Lu, Li-Ming Wu* and Ling Chen*

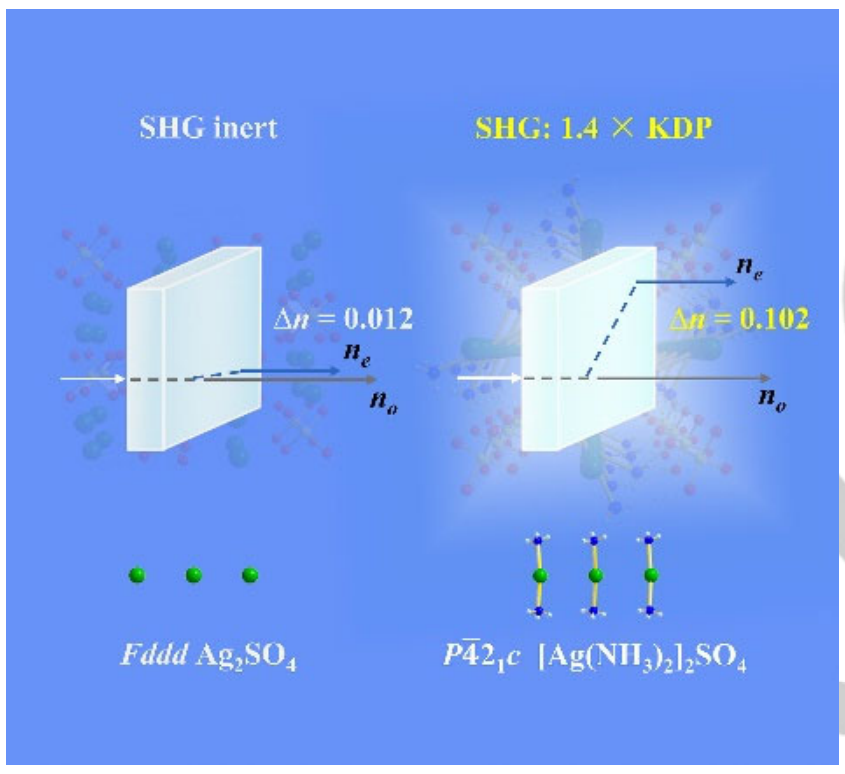

A coordination strategy on the cationic-moiety is proposed for the first time, as demonstrated by the first example, $[Ag(NH_3)_2]_2SO_4$ vs $Ag_2SO_4$, remarkable property enhancement, a strong SHG (1.4 × KDP vs 0 ), a large birefringence ($\Delta n_{cal.}$: 0.102 vs 0.012) are reallized.

# Electronic Supplementary Information

## $[Ag(NH_3)_2]_2SO_4$: A cationic-moeity coordination strategy to design nonlinear optical materials

Yi-Chang Yang, Xin Liu, Jing Lu, Li-Ming Wu* and Ling Chen*

Beijing Key Laboratory of Energy Conversion and Storage Materials
College of Chemistry, Beijing Normal University, Beijing 100875 (P. R. China)

### Table of Contents

## Experimental Procedures

**Synthesis of $[Ag(NH_3)_2]_2SO_4$.** Single crystals of $[Ag(NH_3)_2]_2SO_4$ were grown by slow evaporation of saturated aqueous ammonia solution of $Ag_2SO_4$ at room temperature. $Ag_2SO_4$ (5.0000 g,0.0160 mol, 99.9%) was added to a beaker containing 25.0 ml of ammonia solution (28% $NH_3$), stirred slowly at room temperature until it was completely dissolved, then seal the beaker with a sealing film and make a few small holes to make the solvent evaporate slowly to reduce the contact of the solution with air and prevent the formation of silver carbonate and other impurities. Powder X-ray diffraction (XRD) analysis confirmed the phase purity.

Powder XRD data were collected using a Bruker model D8 Advance powder diffractometer equipped with a Cu Kα radiation source (λ = 1.5418 Å) at room temperature in the 2θ range of 5−70° with a scan step width of 0.02°.

**Single-Crystal Structure Determination.** Although the single crystal structures of $[Ag(NH_3)_2]_2SO_4$ are known,[1] we still collected the single-crystal X-ray diffraction data for further confirmation. A Bruker D8 Quest single-crystal diffractometer, with mirror-monochromatic INCOATEC I*μ*S micro-focus radiation source (50 kV per 1.4 mA), was used to collect the data at room temperature with the aid of APEX III software. The structures were solved and refined with the aid of SHELXS and PLATON softwares. Crystallographic data and structure refinements are given in Table S1, atomic coordinates, equivalent isotropic displacement parameters for $[Ag(NH_3)_2]_2SO_4$ are listed in Table S2, selected bond lengths and angles of the compounds are listed in Table S3.

**Infrared Spectra and UV-vis-NIR Diffuse Reflectance Spectra.** The IR spectra of $[Ag(NH_3)_2]_2SO_4$ were acquired on a Nicolet IS50 FTIR spectrophotometer in the range of 400−4000 $cm^{-1}$ with KBr used as a reference. The UV-vis-NIR diffuse reflectance spectra of $[Ag(NH_3)_2]_2SO_4$ and $Ag_2SO_4$ were collected by a Shimadzu solid Spec-3700DUV spectrophotometer in the wavelength range of 200−1000 nm with a $BaSO_4$ plate as a standard at room temperature.

**Thermal Analysis.** The differential scanning calorimetric (DSC) and thermal gravimetric (TG) data were collected by NETZSCH STA 449 F5 thermal analyzer under $N_2$ atmosphere. The sample was packed in $Al_2O_3$ crucibles and heated from 50 °C to 270 °C or 30 °C to 90 °C at a rate of 1.0 °C/min.

**Second-harmonic Generation.** Powder SHG was measured on polycrystalline samples using the Kurtz and Perry method[2] with Q-switched Nd: YAG lasers at wavelengths of 1064 nm. The polycrystalline samples were ground and sieved into particle size of 25–45, 45–75, 75–109, 109–150 and 150–212 μm, respectively and Commercial KDP was sieved into the corresponding particle size ranges as reference.

**Refractive Index Measurements.** The birefringence of $[Ag(NH_3)_2]_2SO_4$ was measured utilizing the immersion technique[3] with the aid of an FGR-0002 Gem Refractometer on a polished (310)-$[Ag(NH_3)_2]_2SO_4$ crystal plates with a thickness of 0.8 mm. Using sodium yellow light (589.3 nm) as light source. The measured values at different angles by rotating crystal are listed in table S4.

**Theoretical Calculations.** The electronic structures and the linear and nonlinear optical properties of $[Ag(NH_3)_2]_2SO_4$ were calculated by using the pseudopotential method in the VASP[4] package and the density functional theory (DFT).[5] The generalized gradient approximation (GGA)[6] was chosen as the exchange correlation functional, and a plane-wave basis with the projector augmented wave (PAW)[7] potentials were used. The pseudopotentials were used to simulate the ion electron interaction of all constituent elements: Ag $4d^{10}5s^1$, N $2s^22p^3$, H $1s^1$, S $3s^23p^4$, O $2s^22p^4$. A kinetic energy cutoff of 500 eV was chosen with Monkhorst-Pack k-point meshes spanning less than 0.05/$Å^3$ in the Brillouin zone. And then we use the optimized structures to calculate the static self-consistency, the density of state and energy band with a dense 0.02/$Å^3$ k-point spacing mesh. According to the Kramers-Kronig transformation,[8] the real part of the dielectric function $\varepsilon_1(\omega)$, the refractive index $n$ and other linear optical properties can be calculated. Based on the so-called length-gauge formalism derived by Aversa and Sipe,[9] utilizing the specific calculation method invented by Professor Zhang et al,[10] the SHG coefficients were calculated. As the target compound is a ionic crystal, we naturally consider comparing the polarizability anisotropy of $Ag(NH_3)_2^+$ cations and $SO_4^{2-}$ anions to study the sources of birefringence and nonlinear optical properties. At the same time, 0.005 a.u. external field intensity was applied to $Ag(NH_3)_2^+$ cation and $SO_4^{2-}$ anion from *x*-axis, *y*-axis and *z*-axis respectively in order to compare the effect of the same field intensity on the dipole moment of the electron density distribution.

## Figures and Tables

**Table S1.** Crystal data and structural refinement details for $[Ag(NH_3)_2]_2SO_4$

| Compound | $[Ag(NH_3)_2]_2SO_4$ | $Ag_2SO_4$[11] |
|---|---|---|
| Formula weight | 379.94 | 311.8 |
| Crystal system | Tetragonal | orthorgonal |
| Crystal color | colorless | — |
| Space group | $P\bar{4}2_1c$ (No. 114) | Fddd O2 (No.70) |
| *a* (Å) | 8.4372(10) | 5.796 |
| *b* (Å) | 8.4372(10) | 10.2238 |
| *c* (Å) | 6.3980(8) | 12.667 |
| *α* (deg.) | 90 | 90 |
| *β* (deg.) | 90 | 90 |
| *γ* (deg.) | 90 | 90 |
| $V$ (Å$^3$) | 455.45(12) | 750.6 |
| $Z$ | 2 | 8 |
| $D_c$ (g·cm$^{-3}$) | 2.770 | 5.52 |
| $\mu$ (mm$^{-1}$) | 4.509 | — |
| F(000) | 364.0 | — |
| GOOF on $F^2$ | 1.224 | — |
| $R_1$, $wR_2$ ($I > 2\sigma(I)$)$^a$ | 0.0213, 0.0552 | — |
| $R_1$, $wR_2$ (all data)$^a$ | 0.0220, 0.0557 | — |

$^aR_1 = \mathbf{\Sigma}||F_o| - |F_c||/\mathbf{\Sigma}|F_o|$ and $wR_2 = [\mathbf{\Sigma}w(F_o^2 - F_c^2)^2/\mathbf{\Sigma}w(F_o^2)^2]^{1/2}$ for $F_o^2 > 2\sigma(F_o^2)$

**Table S2.** Atomic coordinates and equivalent isotropic displacement parameters of $[Ag(NH_3)_2]_2SO_4$. $U_{eq}$ is defined as one-third of the trace of the orthogonalized $U_{ij}$ tensor.

| Atom | Wyckoff | *x* | *y* | *z* | $U_{eq}$ (Å$^2$) |
|---|---|---|---|---|---|
| Ag1 | 4*d* | 0 | 0.5000 | 0.49199(5) | 0.0340(3) |
| S1 | 2*a* | 0.5000 | 0.5000 | 0.5000 | 0.0199(4) |
| N1 | 8*e* | 0.1090(3) | 0.7244(3) | 0.5082(5) | 0.0314(6) |
| O1 | 8*e* | 0.4406(3) | 0.6296(3) | 0.6319(4) | 0.0391(6) |
| H1 | 8*e* | 0.067(5) | 0.774(5) | 0.602(6) | 0.046(11) |
| H2 | 8*e* | 0.092(5) | 0.779(5) | 0.401(7) | 0.043(10) |
| H3 | 8*e* | 0.224(6) | 0.722(5) | 0.532(10) | 0.035(13) |

**Table S3.** Selected bond lengths (Å) and angles (deg.) of $[Ag(NH_3)_2]_2SO_4$.

| | | | |
|---|---|---|---|
| Ag(1)-Ag(1)$^{\#1}$ | 3.1990(4) | S(1)-O(1) | 1.469(2) |
| Ag(1)-Ag(1)$^{\#2}$ | 3.1990(4) | S(1)-O(1)$^{\#4}$ | 1.469(2) |
| Ag(1)-N(1) | 2.107(3) | S(1)-O(1)$^{\#5}$ | 1.469(2) |
| Ag(1)-N(1)$^{\#3}$ | 2.107(3) | S(1)-O(1)$^{\#6}$ | 1.469(2) |
| N(1)-H(1) | 0.815 | N(1)-H(3) | 0.985 |
| N(1)-H(2) | 0.834 | | |
| Ag(1)$^{\#1}$-Ag(1)-Ag(1)$^{\#2}$ | 180.0 | O(1)$^{\#4}$-S(1)-O(1)$^{\#5}$ | 109.9(2) |
| N(1)$^{\#3}$-Ag(1)-Ag(1)$^{\#1}$ | 87.18(8) | O(1)$^{\#4}$-S(1)-O(1)$^{\#6}$ | 109.25(10) |
| N(1)$^{\#3}$-Ag(1)-Ag(1)$^{\#2}$ | 92.82(8) | O(1)$^{\#5}$-S(1)-O(1)$^{\#6}$ | 109.25(10) |
| N(1)-Ag(1)-Ag(1)$^{\#2}$ | 92.82(8) | O(1)$^{\#6}$-S(1)-O(1) | 109.9(2) |
| N(1)-Ag(1) -Ag(1)$^{\#1}$ | 87.18(8) | O(1)$^{\#5}$-S(1)-O(1) | 109.25(10) |
| N(1)$^{\#3}$-Ag(1)-N(1) | 174.35(17) | O(1)$^{\#4}$-S(1)-O(1) | 109.25(10) |
| Ag(1)-N(1)-H(2) | 112(3) | Ag(1)-N(1)-H(3) | 115(3) |
| Ag(1)-N(1)-H(1) | 108(3) | H(2)-N(1)-H(3) | 108(5) |
| H(2)-N(1)-H(1) | 104(4) | H(3)-N(1)-H(1) | 109(4) |

Symmetry transformations used to generate equivalent atoms:
#1 1/2-Y, 1/2-X, 1/2+Z; #2 1/2-Y, 1/2-X, -1/2+Z; #3 -X, 1-Y, +Z; #4 1-Y, +X, 1-Z; #5 +Y, 1-X, 1-Z; #6 1-X, 1-Y, +Z.

Table S4. The charge density of $[Ag(NH_3)_2]^+$ and $SO_4^{2-}$.

| | | $Ag(NH_3)_2^+$ | $SO_4^{2-}$ |
|---|---|---|---|
| polarizability | XX | 33.05 | 33.49 |
| | XY | -0.00086 | 0 |
| | YY | 45.92 | 33.48 |
| | XZ | 0 | -0.0013 |
| | YZ | 0 | 0 |
| | ZZ | 33.10 | 33.48 |
| Isotropic average polarizability | | 37.36 | 33.48 |
| Polarizability anisotropy | | 12.84 | 0.0093 |
| static dipole moment/D | X | 0 | -0.00051 |
| | Y | 0 | 0 |
| | Z | -0.12 | 0.0020 |
| | Tot | 0.12 | 0.0020 |
| *x*+0.005 a.u. dipole moment/D | X | -0.42 | -0.43 |
| | Y | 0 | 0 |
| | Z | -0.31 | 0.0052 |
| | Tot | 0.52 | 0.43 |
| *y*+0.005 a.u. dipole moment/D | X | 0 | -0.0012 |
| | Y | -0.58 | -0.43 |
| | Z | -0.31 | 0.0052 |
| | Tot | 0.66 | 0.43 |
| *z*+0.005 a.u. dipole moment/D | X | 0 | -0.0013 |
| | Y | 0 | 0 |
| | Z | -0.73 | -0.42 |
| | Tot | 0.73 | 0.42 |

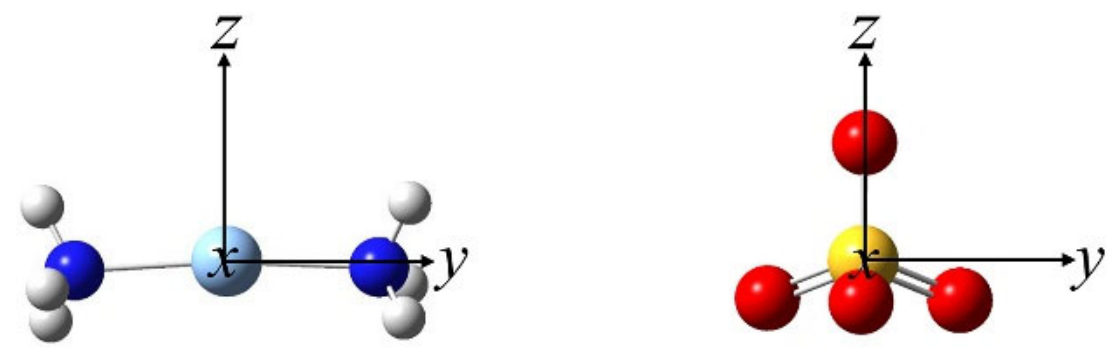

Figure S1. Geometry of isolated $[Ag(NH_3)_2]^+$ and $SO_4^{2-}$ and the direction of applied electric field.

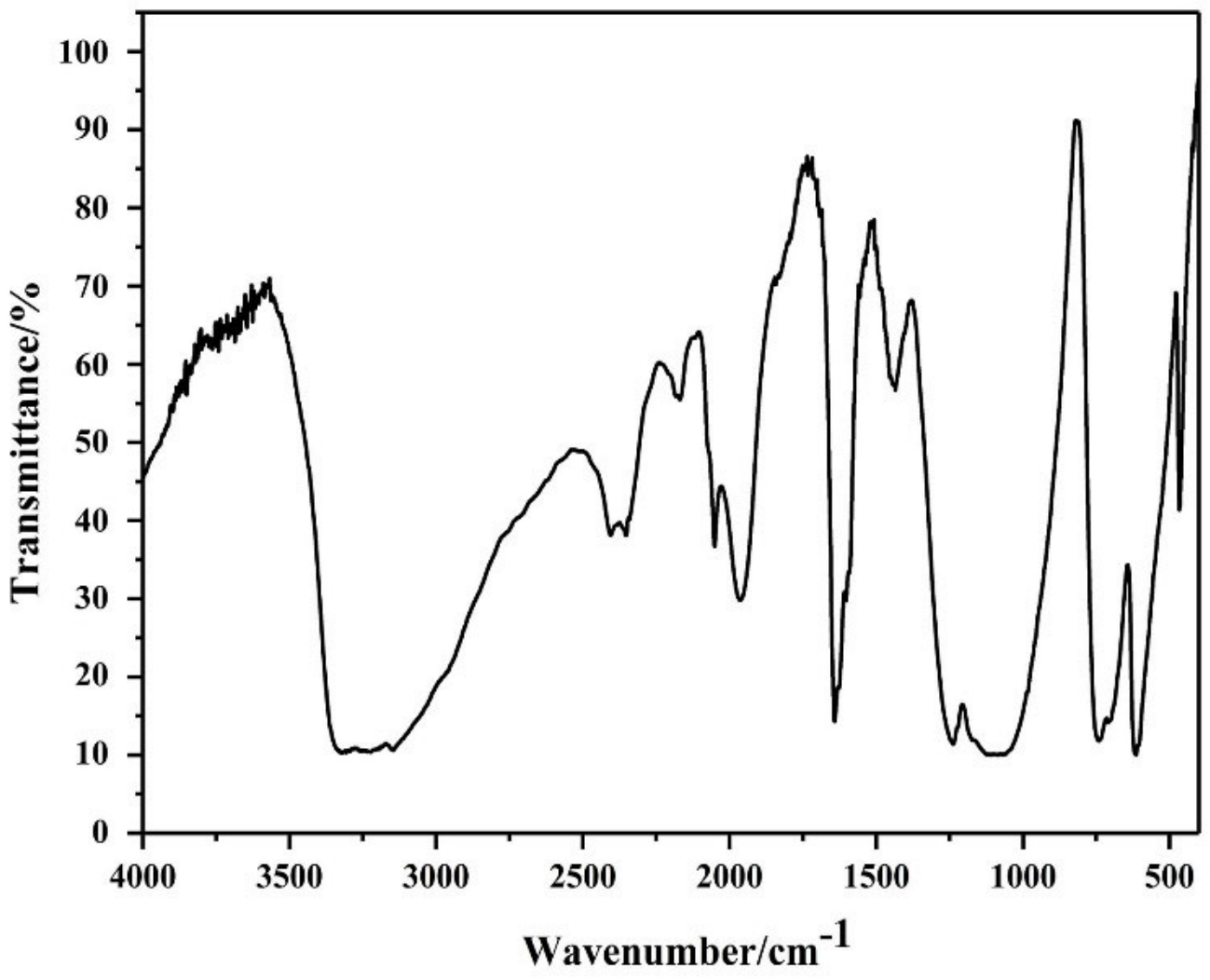

**Figure S2.** IR spectrum for $[Ag(NH_3)_2]_2SO_4$.

Table S5. Assignment of the absorption bands observed in the IR spectra for $[Ag(NH_3)_2]_2SO_4$.

| Assignment | $\upsilon_{as}(NH_3)$ | $\upsilon_s(NH_3)$ | $\upsilon_{as}(AgN_2)$ | $\delta_{as}(NH_3)$ | $\delta_s(NH_3)$ | $\rho_r(NH_3)$ |
|---|---|---|---|---|---|---|
| $[Ag(NH_3)_2]_2SO_4$ | 3320 | 3150 | 476 | 1641 | 1243 | 743/702 |

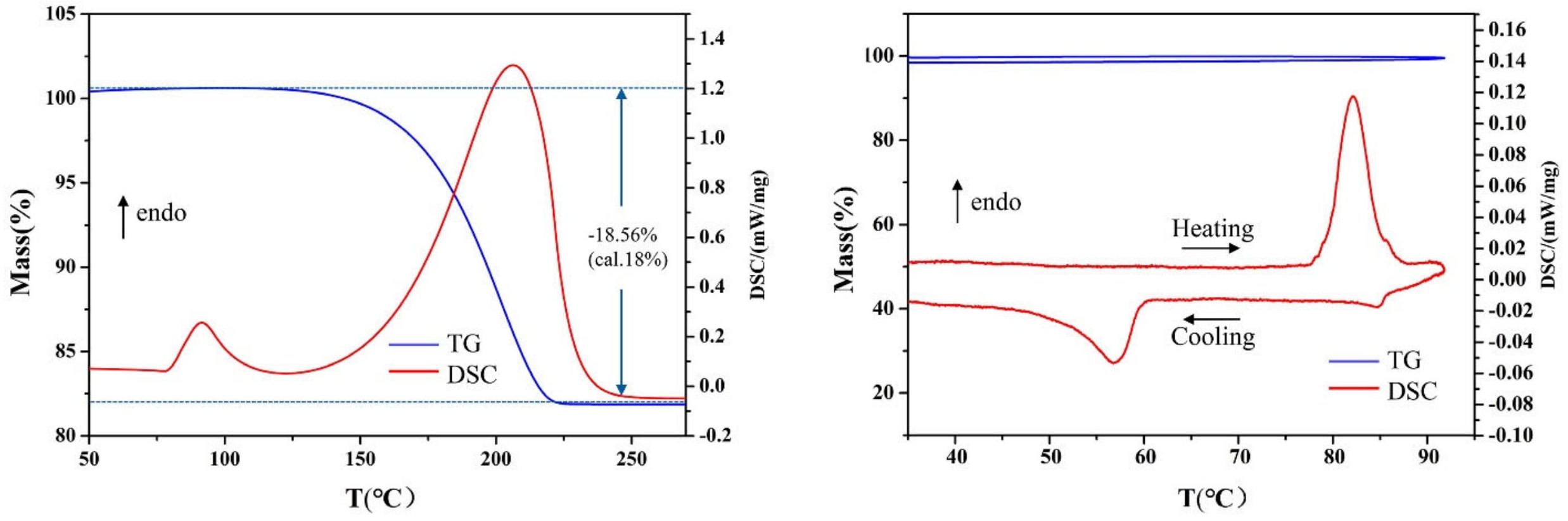

**Figure S3.** TG and DSC curves of $[Ag(NH_3)_2]_2SO_4$, the temperature range is (a) 50 °C to 270 °C and (b) 30 °C to 90 °C. In order to confirm the 1st order phase transition conjecture, as the crystalline powder sample is heated to 90 °C and then cooled to room temperature (Figure S3b), the sample is still powdery without melting. The power sample after DSC testing is analyzed by XRD (Figure S4), the main phase is still $[Ag(NH_3)_2]_2SO_4$ which indicates that the endothermic peak around 83 °C and exothermic peak around 57 °C corresponds to the phase transformation.

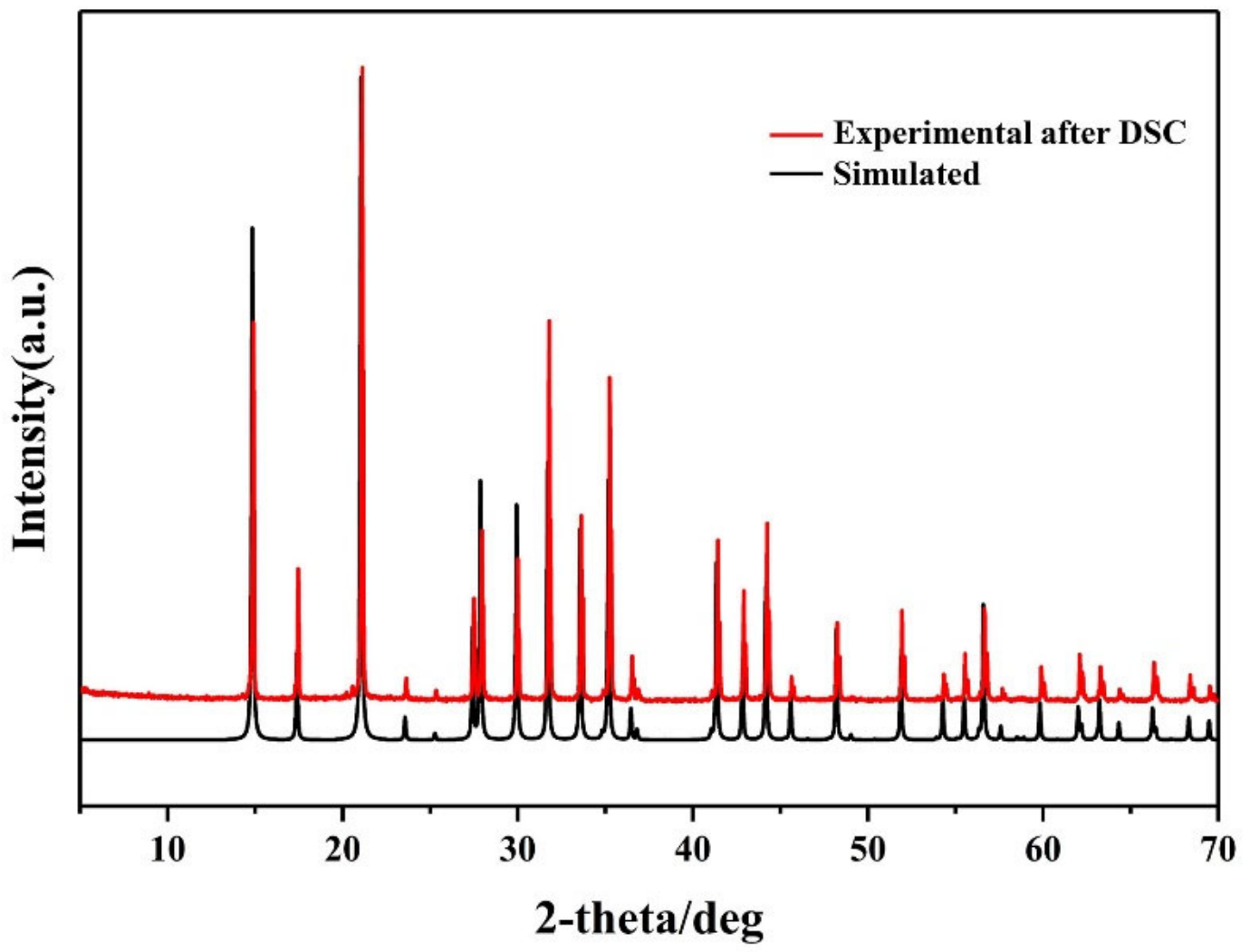

Figure S4. Experimental PXRD of $[Ag(NH_3)_2]_2SO_4$ after DSC testing up to 90 °C and simulated PXRD patterns of $[Ag(NH_3)_2]_2SO_4$.

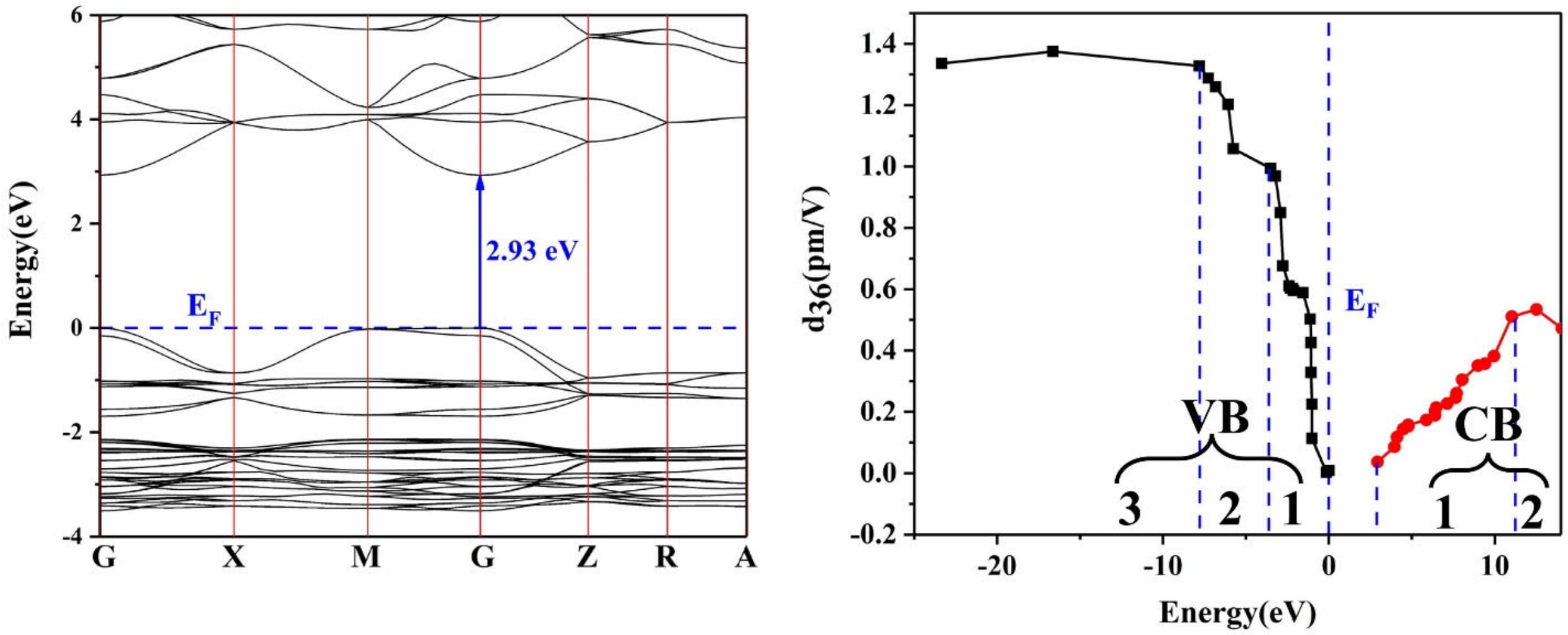

**Figure S5.** (a) Calculated band structures, (b) Cutoff-energy-dependent static SHG coefficient of $[Ag(NH_3)_2]_2SO_4$.

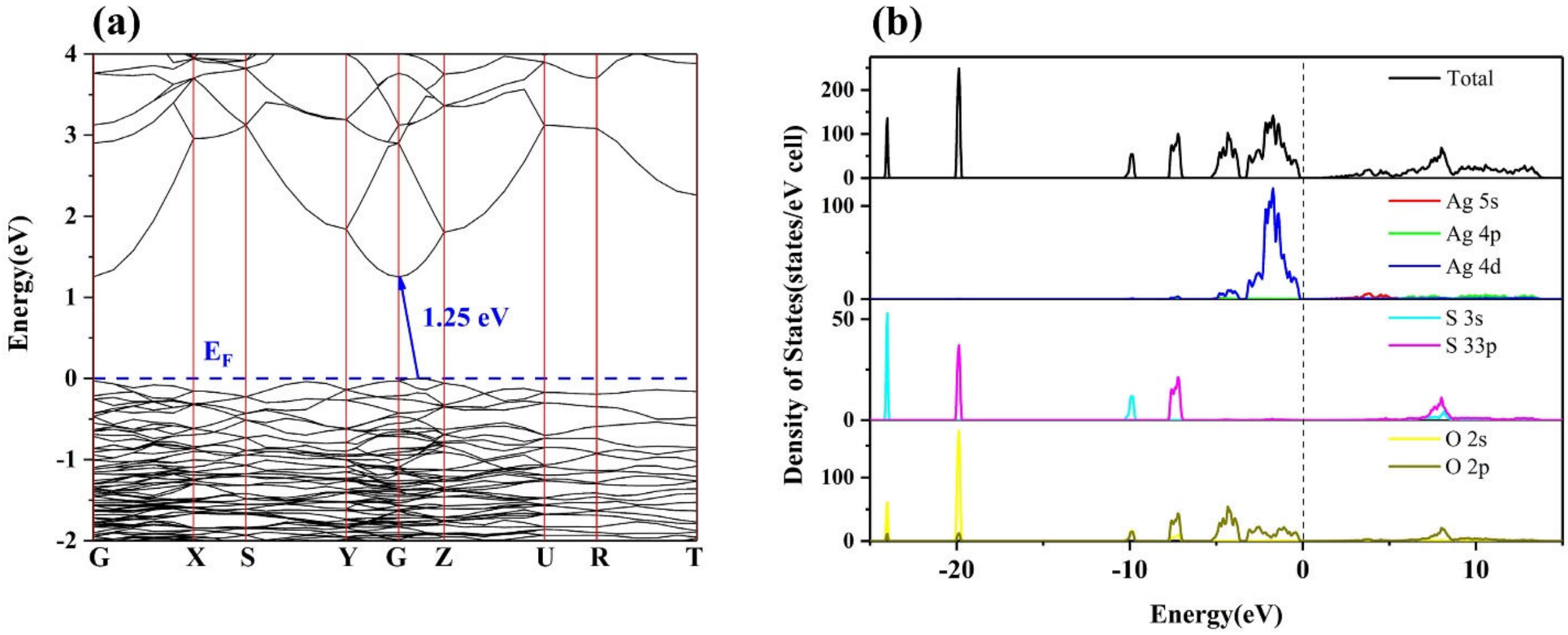

**Figure S6.** (a) Calculated band structures and (b) Total and partial densities of states of $Ag_2SO_4$.

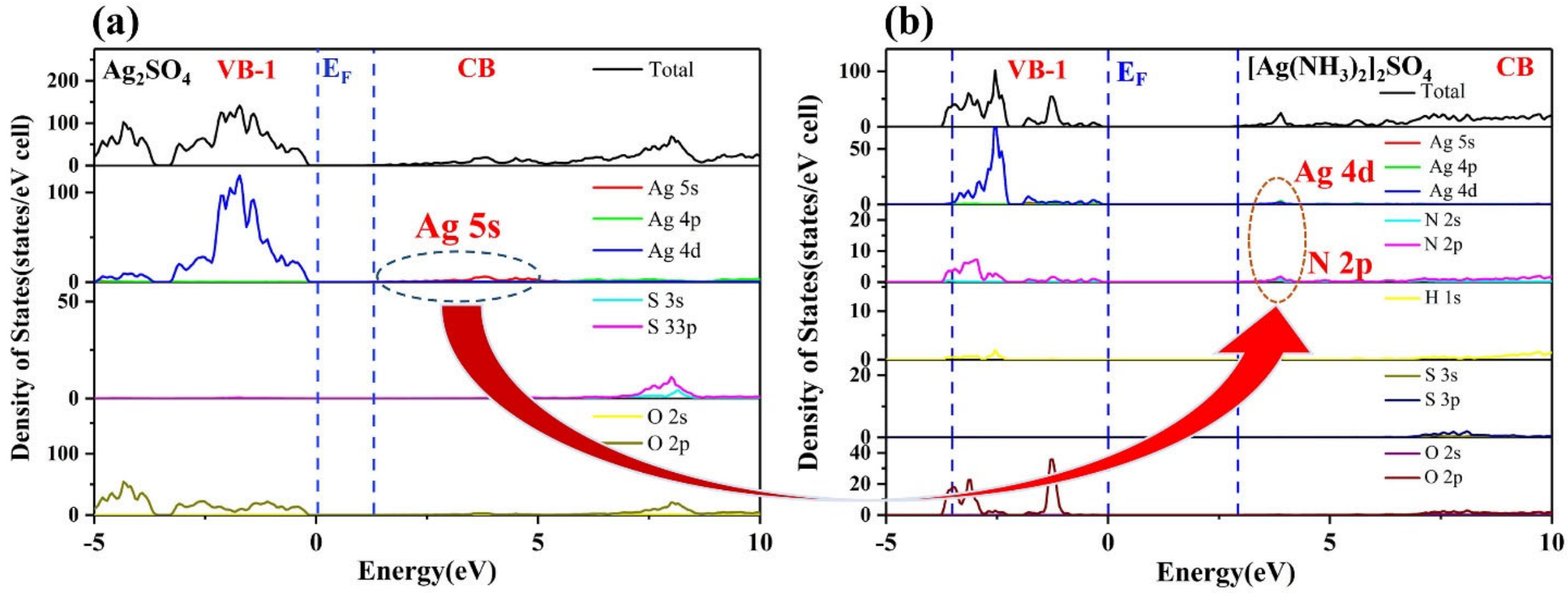

Figure S7. Total and partial densities of states of $Ag_2SO_4$ and $[Ag(NH_3)_2]_2SO_4$ near the Fermi level.

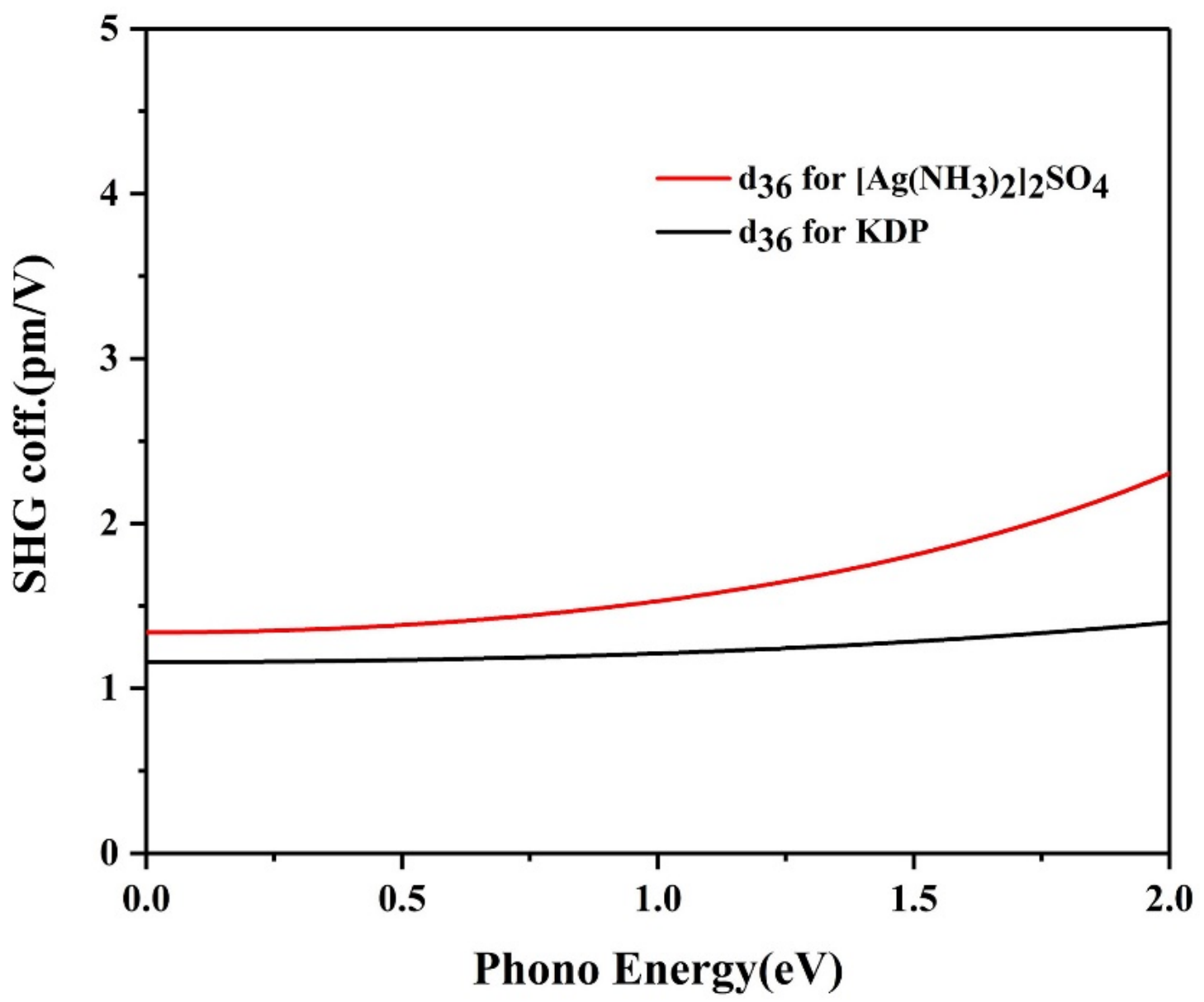

**Figure S8.** Calculated frequency-dependent second-harmonic generation coefficients for $[Ag(NH_3)_2]_2SO_4$ and KDP.

Table S6. Experimental refractive index measured on Single Crystal plates at 589.3 nm. ($n_{max}$ and $n_{min}$ are shown in red).

| Single Crystal $[Ag(NH_3)_2]_2SO_4$ - (310) plate | | | | | | |
|---|---|---|---|---|---|---|
| ***angle**** | **0°** | **30°** | **60°** | **90°** | **120°** | **150°** |
| ***n*** | 1.612<br>1.692 | 1.609<br>1.669 | 1.618<br>1.635 | 1.605 | 1.613<br>1.629 | 1.608<br>1.667 |
| ***angle**** | **180°** | **210°** | **240°** | **270°** | **300°** | **330°** |
| ***n*** | 1.619<br>1.693 | 1.622<br>1.685 | 1.629<br>1.646 | 1.622 | 1.619<br>1.638 | 1.614<br>1.662 |

*angle: the definition is shown in Figure S5.

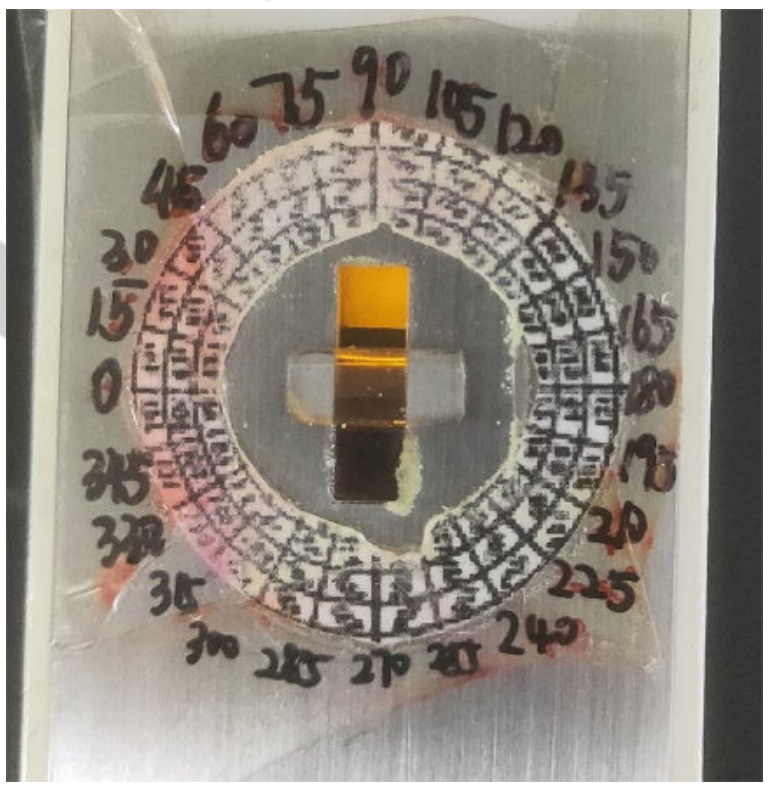

Figure S9. the definition of the rotation angle during the refractive index measurement.

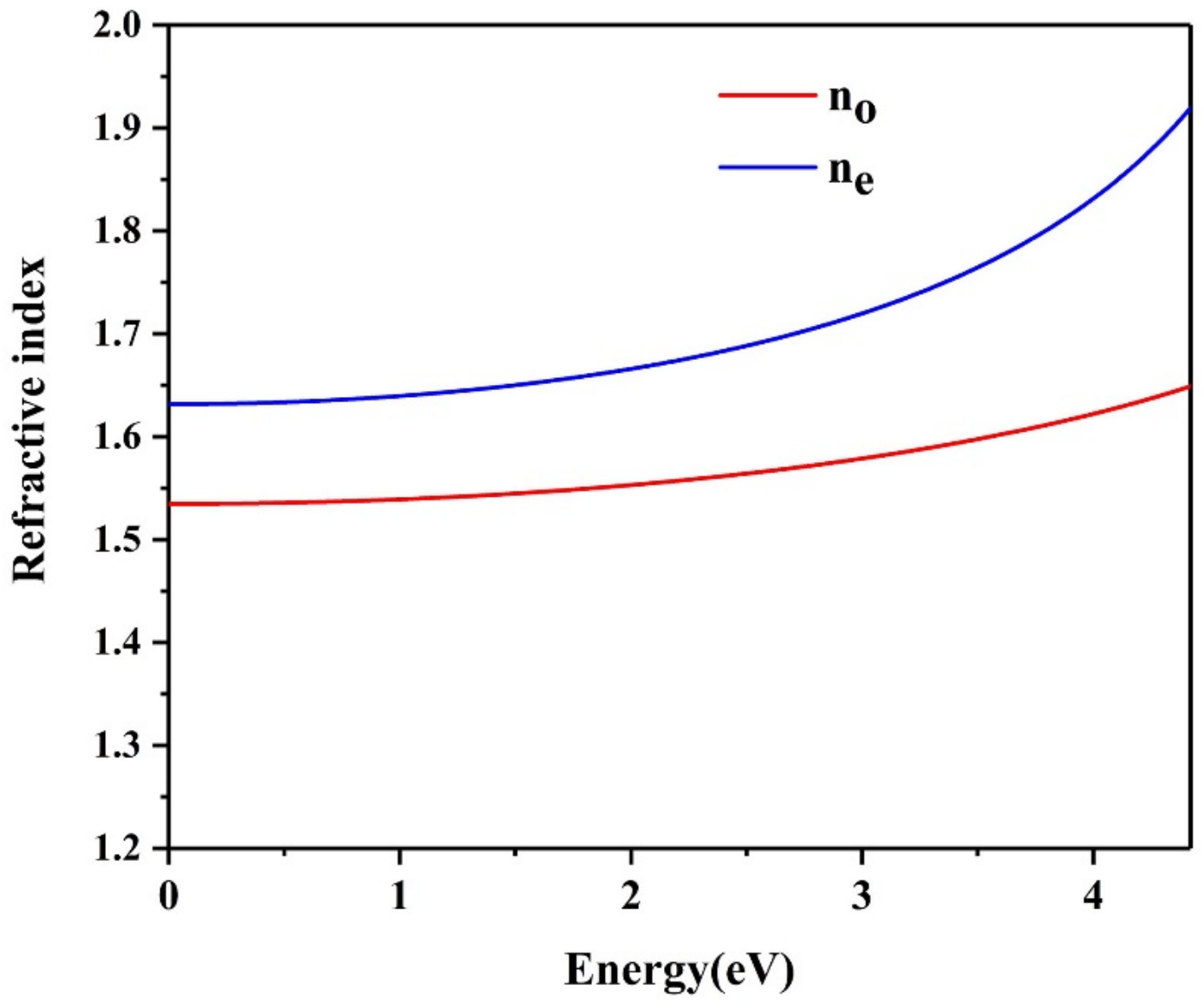

**Figure S10.** Calculated linear refractive indices for $[Ag(NH_3)_2]_2SO_4$.

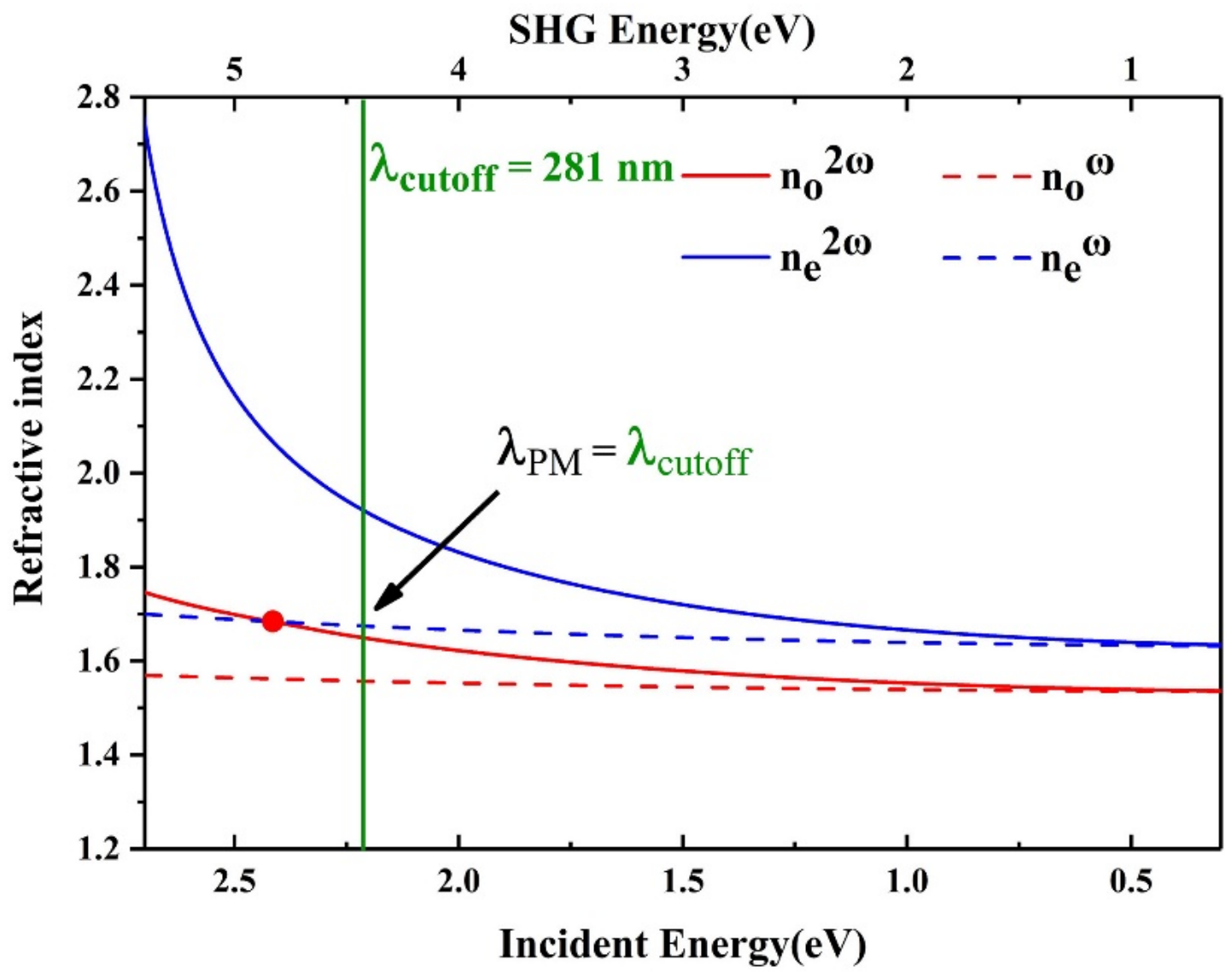

**Figure S11.** The calculated refractive index dispersion curves of $[Ag(NH_3)_2]_2SO_4$ indicating the shortest SHG λPM at 281 nm.

## References:

[1] U. Zachwieja, H. Jacobs, *Z. Kristallogr.* **1992**, *201*, 207-212.
[2] S. K. Kurtz, T. T. Perry, *J. Appl. Phys.* **1968**, *39*, 3798-3813.
[3] C. T. Chen, Y. B. Wang, Y. N. Xia, B. C. Wu, D. Y. Tang, K. C. Wu, W. R. Zeng, L. H. Yu, L. F. Mei, *J. Appl. Phys.* **1995**, *77*, 2268-2272.
[4] G. Kresse, J. Furthmuller, *Phys. Rev. B* **1996**, *54*, 11169-11186.
[5] W. Kohn, *Rev. Mod. Phys.* **1999**, *71*, 1253-1266.
[6] J. P. Perdew, Y. Wang, *Phys. Rev. B* **1992**, *45*, 13244-13249.
[7] G. Kresse, D. Joubert, *Phys. Rev. B* **1999**, *59*, 1758-1775.
[8] a) S. Laksari, A. Chahed, N. Abbouni, O. Benhelal, B. Abbar, *Comput. Mater. Sci.* **2006**, *38*, 223-230; b) S. D. Mo, W. Y. Ching, *Phys. Rev. B* **1995**, *51*, 13023-13032.
[9] a) C. Aversa, J. E. Sipe, *Phys. Rev. B* **1995**, *52*, 14636-14645; b) S. N. Rashkeev, W. R. L. Lambrecht, B. Segall, *Phys. Rev. B* **1998**, *57*, 3905-3919.
[10] Z. Fang, J. Lin, R. Liu, P. Liu, Y. Li, X. Huang, K. Ding, L. Ning, Y. Zhang, *Crystengcomm* **2014**, *16*, 10569-10580.
[11] N. E. Brese, M. Okeeffe, B. L. Ramakrishna, R. B. Vondreele, *J. Solid State Chem.* **1990**, *89*, 184-190.